\documentclass[12pt,preprint]{aastex}

\usepackage{amsmath,natbib,graphicx}
\usepackage{epsf}
\usepackage{fancyvrb}
\usepackage{epsfig}
\usepackage{color}
\usepackage{lscape}
\DeclareGraphicsExtensions{.jpg,.pdf,.png,.eps,.ps}

\begin{document}
\VerbatimFootnotes

\title{Decrease in hysteresis of planetary climate for planets with long solar days}
\shorttitle{Solar Day and Snowball Bifurcation}


\author{Dorian~S.~Abbot\altaffilmark{1}, Jonah Bloch-Johnson\altaffilmark{1}, Jade Checlair\altaffilmark{1}, Navah X. Farahat\altaffilmark{1}, R.~J. Graham\altaffilmark{1}, David Plotkin\altaffilmark{1}, Predrag Popovic\altaffilmark{1}, and Francisco Spaulding-Astudillo\altaffilmark{1}}
\altaffiltext{1}{Department of the Geophysical Sciences, University of
  Chicago, 5734 South Ellis Avenue, Chicago, IL 60637}
\shortauthors{Abbot}

\email{abbot@uchicago.edu}


\begin{abstract}
  The ice-albedo feedback on rapidly-rotating terrestrial planets in
  the habitable zone can lead to abrupt transitions (bifurcations)
  between a warm and a snowball (ice-covered) state, bistability
  between these states, and hysteresis in planetary climate. This is
  important for planetary habitability because snowball events may
  trigger rises in the complexity of life, but could also endanger
  complex life that already exists. Recent work has shown that planets
  tidally locked in synchronous rotation states will transition
  smoothly into the snowball state rather than experiencing
  bifurcations. Here we investigate the structure of snowball
  bifurcations on planets that are tidally influenced, but not
  synchronously rotating, so that they experience long solar days. We
  use PlaSIM, an intermediate-complexity global climate model, with a
  thermodynamic mixed layer ocean and the Sun's spectrum. We find that
  the amount of hysteresis (range in stellar flux for which there is
  bistability in climate) is significantly reduced for solar days with
  lengths of tens of Earth days, and disappears for solar days of
  hundreds of Earth days. These results suggest that tidally
  influenced planets orbiting M and K-stars that are not synchronously
  rotating could have much less hysteresis associated with the
  snowball bifurcations than they would if they were rapidly rotating.
  This implies that the amount of time it takes them to escape a
  snowball state via CO$_2$ outgassing would be greatly reduced, as
  would the period of cycling between the warm and snowball state if
  they have a low CO$_2$ outgassing rate.
\end{abstract}

\bigskip\bigskip

\section{Introduction}
\label{sec:introduction}

Earth has experienced a few episodes of global, or very near-global,
ice coverage that are called ``snowball Earth'' events
\citep{Kirschvink92,Hoffman98}. These events may have been critical
for increasing the complexity of life, both directly through forcing
evolutionary innovations \citep{Kirschvink92,Hoffman98,Hoffman02} and
indirectly by increasing atmospheric oxygen
\citep{Hoffman02,laakso2014regulation,laakso2017}. The snowball events
on Earth were isolated and lasted for millions to tens of millions of
years \citep{rooney2015cryogenian}. Habitable zone exoplanets with low
CO$_2$ outgassing rates, however, could experience perpetual cycling
between snowball and warm states, with most of the time spent in
snowball conditions
\citep{Tajika2007,Kadoya:2014kd,Menou2015,haqq2016limit,batalha2016climate,abbot-2016,paradise2017}.
Understanding how snowball episodes might function on Earth-like
exoplanets is therefore critical for planetary habitability.

\begin{figure}[h!]
\begin{center}
\epsfig{file=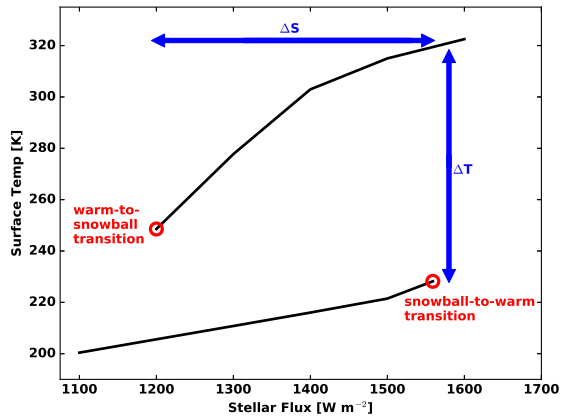, width=0.5\textwidth} 
\end{center}
\caption{This is a plot of a closed hysteresis loop over the allowable
  states of global mean surface temperature as a function of stellar
  flux for an idealized rapidly rotating planet (black lines). The two
  bifurcations in the system are represented by red circles and
  labeled ``warm-to-snowball transition'' and ``snowball-to-warm
  transition.'' The range of stellar flux over which the planet is
  bistable ($\Delta$S) and the increase in global mean surface
  temperature associated with the snowball-to-warm transition
  ($\Delta$T) are represented by blue double arrows.}
\label{fig:schematic}
\end{figure}

The ice-albedo feedback occurs when cooling causes the replacement of
low-albedo ocean with high-albedo ice, which leads to more cooling and
ice formation. As a result of the ice-albedo feedback, rapidly
rotating planets exhibit a bifurcation, or jump, in climate, into and
out of snowball events \citep{Budyko-1969:effect,Sellers-1969:ae}.
These bifurcations are associated with bistability for a range of
radiative forcing, as well as hysteresis. This is shown schematically
in Fig.~\ref{fig:schematic}. If we start in the warm state and
decrease the stellar flux (or equivalently CO$_2$), eventually the
climate transitions into the snowball state through a bifurcation at
some critical value of the stellar flux. We will refer to this process
as the ``warm-to-snowball transition.'' If we then increase the
stellar flux, the climate remains in the snowball state even at
stellar fluxes where it was previously in the warm state, which
reflects the bistability of the system. The ``snowball-to-warm
transition'' occurs when the stellar flux is increased enough for a
planet in the snowball state to jump into the warm state. The path
dependence of the system that results from bistability is referred to
as hysteresis. We can quantify the amount of hysteresis in the system
through the range of stellar flux over which the planet is bistable
($\Delta$S) and the difference between the warm and snowball states
through the increase in global mean surface temperature associated
with the snowball-to-warm transition ($\Delta$T)
(Fig.~\ref{fig:schematic}).

Tidal locking, including into a synchronous 1:1 spin-orbit state,
should be common for planets in the habitable zones of the many M and
K stars in the solar neighborhood that will be prime targets for
future observational missions \citep{Kasting93,barnes2016tidal}. For
this reason \citet{checlair2017} investigated the effect of
synchronous tidal locking on the snowball state. They used PlaSim, an
intermediate-complexity global climate model (GCM), to show that
planets in synchronous rotation states are unlikely to experience
bifurcations and hysteresis associated with the snowball state.
Instead, they should transition smoothly from the warm to the snowball
state. \citet{checlair2017} explained the loss of the snowball
bifurcation as resulting directly from the change in the pattern of
incoming radiation from the central star, in particular the strong
increase in incoming radiation as the substellar point is approached.
The loss of hysteresis in planetary climate is important because it
means that a synchronously rotating planet with an active carbon cycle
\citep{Walker-Hays-Kasting-1981:negative} could not remain in a
snowball state. If such a planet happened into a snowball state,
weathering would be greatly reduced or zero, and without hysteresis
volcanic outgassing of CO$_2$ would immediately warm the planet enough
to expose open ocean at the substellar point.

Tidally influenced planets will not all be synchronously rotating. For
high enough eccentricities they will be caught in higher order spin
states \citep{barnes2016tidal} and even in a circular orbit they can
avoid synchronous rotation because of atmospheric tides
\citep{leconte2015asynchronous}. This raises the question of what
happens to the snowball bifurcation on a planet that is not
synchronously rotating, but has a very long solar day (the time from
local noon to the next local noon). \citet{boschi2013changes} and
\citet{lucarini2013habitability} did some simulations to investigate
this using PlaSim with no continents, an ocean with a depth of 50~m,
and no ocean heat transport, and found that hysteresis is lost for
planets with a solar day greater than about 180 Earth days. In a
related study, \citet{salameh2017role} investigated the stellar flux
that leads to global glaciation as a function of planetary rotation
rate using ECHAM6, a more sophisticated atmospheric GCM. They found
that decreasing the rotation rate has a complicated, non-monotonic
effect on the stellar flux at which the transition to global glacation
occurs, with the transition ultimately occurring at a much lower
stellar flux for a synchronously rotating planet.

The purpose of this paper is to study in more depth the transition
from a planet with a snowball bifurcation and hysteresis for a short
solar day to one without them for a long solar day. We will be
particularly interested in determining the critical solar day length
at which this change in behavior occurs, understanding in detail what
physically leads to this transition, and outlining the implications
for planets that will be observable in the near future. In order to
isolate the effect of the solar day, we will not consider changes in
the planetary rotation rate \citep{salameh2017role} or the reduction
in the ice-albedo feedback for redder stars
\citep{Joshi:2012hu,shields2013effect,shields2014spectrum}.

We will use the intermediate complexity GCM PlaSim in this study, run
with no continents and a thermodynamic mixed layer ocean with no heat
transport and a fixed depth. This model has a lower resolution and
somewhat less sophisticated physical parameterizations than the GCMs
used to forecast climate change. Its main advantage is that it is
numerically efficient so that we can perform large numbers of
simulations to determine accurately the locations of bifurcations in a
variety of situations. More complex 3D atmospheric GCMs
\citep{Abbot-Pierrehumbert-2009:mudball,abbot12-snowball-clouds,Voigt2012,voigt2013dynamics,abbot13-snowball-circulation,Leconte:2013gv,leconte2013increased,yang2013,shields2013effect,shields2014spectrum,yang2014,Wolf:2014,wolf2015evolution,way2015exploring,kopparapu2016inner,salameh2017role,wolf2017assessing,wolf2017constraints,haqq2017demarcating}
and models that include ocean
\citep{yang2011a,yang2011b,yang2012,voigt11,voigt12-dynamics,hu2014role,way2015exploring,cullum2016importance}
and ice dynamics
\citep{Goodman03,Pollard05,Goodman06,li2011sea,Tziperman2012,abbot13-snowball-circulation,pollard2017snowball}
have been applied to snowball and exoplanet climate before. It will be
important to check our results with models including more processes in
the future, insofar as this is possible given numerical constraints.
Nevertheless this work is an important first step and outlines
physical processes that are likely to be active even in these more
sophisticated models.

\section{Methods}
\label{sec:methods}

\subsection{Model}
\label{sec:model}

We use PlaSim \citep{fraedrich2005planet}, the same
intermediate-complexity 3D global climate model (GCM) used by
\citet{checlair2017}, \citet{boschi2013changes}, and
\citet{lucarini2013habitability}. PlaSim solves the primitive
equations for atmospheric dynamics and has schemes to calculate
radiative transfer, convection and clouds, thermodynamic sea ice, and
land-atmosphere interactions. We run the model at T21 horizontal
resolution ($5.625^\circ \times 5.625^\circ$) with 10 vertical levels.
We use the modern solar spectral distribution, so we do not take into
account the stellar effect on the ice-albedo feedback
\citep{Joshi:2012hu,shields2013effect,shields2014spectrum}. We use an
atmospheric CO$_2$ of 360 ppm and set other trace greenhouse gases to
zero. PlaSim calculates H$_2$O prognostically and includes its
radiative effects. The surface boundary is a thermodynamic mixed layer
(slab) ocean with zero imposed ocean heat transport and a mixed layer
depth that we vary. A mixed layer ocean is essentially a motionless
layer of water that provides the surface with a heat capacity
proportional to its depth. We use an aquaplanet configuration, with no
continents. We set the eccentricity and obliquity to zero.


We use the same basic methodology as \citet{checlair2017}. We first
perform simulations to obtain equilibrated climates where the surface
is either ice-free (henceforth warm start) or 100\% ice-covered
(henceforth cold start). For a particular parameter choice, we run
simulations starting from both warm and cold starts at a variety of
different stellar fluxes. We then use a bisection algorithm to find
the stellar fluxes at which bifurcations occur to within
0.5~W~m$^{-2}$. We run all simulations until they reach
top-of-atmosphere and surface energy balance, typically 60--100 Earth
years. We average all displayed variables over 10 Earth years after a
simulation has converged. We also performed test simulations along the
warm branch of solutions in which we decreased the stellar flux in
successive increments of 5~W~m$^{-2}$ for each simulation and
restarted from the previous converged simulation. We obtained the same
results using this method as we did using the standard warm start
methodology.
 
\subsection{Changing the Solar Day}
\label{sec:solarday}

Following \citet{boschi2013changes} and
\citet{lucarini2013habitability}, we consider longer solar days by
decreasing the length of the year while keeping the planetary rotation
rate fixed. This section describes how we modified
the radiation scheme to accomplish this.

$T_{orb}$, the orbital period, is the amount of time it takes the
planet to orbit its star. The sidereal day,
$T_{sid}$, is the amount of time it takes for the planet to do
one rotation around its own axis relative to the distant stars. This
means that the number of rotations per orbital period is
\begin{equation}
N_{sid}=\frac{T_{orb}}{T_{sid}}.
\label{eq:nsid}
\end{equation}
We will take the sidereal day to be constant, but allow $N_{sid}$ to
change. The solar day, $T_{sol}$, is the amount time from local noon
to the next local noon on the planet's surface. The number of solar
days per orbital period is
\begin{equation}
N_{sol}=\frac{T_{orb}}{T_{sol}}.
\label{eq:nsol}
\end{equation}
For a prograde orbit, we must have
\begin{equation}
N_{sid}=1+N_{sol},
\label{eq:nsid-nsol}
\end{equation}
because the planet orbiting once around the central star cancels out
one solar day that the planet would have experienced due to rotation
alone. Finally, define $\theta$ as the longitude of the planet where
the central star is directly overhead in radians. Assuming a constant
planetary rotation rate and constant motion of the planet around the
star (a circular orbit), we find that
\begin{equation}
\theta = 2\pi \frac{t}{T_{sol}},
\label{eq:theta}
\end{equation}
where $t$ is the time, which we will keep track of in Earth minutes
(60 seconds).

Now let us suppose that we wish to consistently change $T_{orb}$ and
$T_{sol}$ while keeping $T_{sid}$ constant. Suppose that we want to
consider a low-order orbital resonance. In this case $N_{sid}$ defines
the resonance, so that, for example, if we consider a 2:1 spin-orbit
resonance, $N_{sid}=2$. Given that we are keeping the rotation rate
(and therefore $T_{sid}$) constant, this would mean a very short
orbital period. We can solve
Equations~(\ref{eq:nsid}-\ref{eq:nsid-nsol}) for the solar day as follows
\begin{equation}
T_{sol}=T_{sid}\frac{N_{sid}}{N_{sid}-1}=\frac{T_{orb}}{N_{sid}-1}.
\label{eq:tsol}
\end{equation}
Now we can combine Equations~(\ref{eq:theta})~and~(\ref{eq:tsol}) to
solve for $\theta$
\begin{equation}
\theta = 2\pi \frac{t}{T_{orb}} (N_{sid}-1)
  =2\pi \frac{t}{T_{sid}} \frac{N_{sid}-1}{N_{sid}},
\label{eq:theta2}
\end{equation}
where it is useful to write $\theta$ this way since we are assuming
that $T_{sid}$ is constant. Now, let us suppose that the planet we are
considering originally, before we change $T_{orb}$, is an idealized
version of modern Earth with an orbital period of 360 solar days, each
lasting $24 \times 60=1440$ minutes. This means the planet has 361
sidereal days, so that 
\begin{equation}
T_{sid}=\frac{360}{361} (1440\ \min).
\label{eq:tsid}
\end{equation}
We are taking $T_{sid}$ to be constant, so we can substitute
Equation~(\ref{eq:tsid}) into Equation~(\ref{eq:theta2}) and find that
\begin{equation}
\theta = 2\pi \frac{t}{(1440\ \min)}
  \frac{361}{360}\frac{N_{sid}-1}{N_{sid}}.
\label{eq:theta3}
\end{equation}
The radiative scheme in PlaSim uses $\theta$ directly, so we can now
specify the position of the central star for different values of
$N_{sid}$ to the code. Using Equation~(\ref{eq:tsol}), this is
equivalent to specifying the position of the central star for
different values of $T_{sol}$.

Before proceeding, let's consider the asymptotic limits of
Equation~(\ref{eq:theta3}). If the planet is tidally locked, then
$N_{sid}=1$, so that by Equation~(\ref{eq:tsol}), $T_{sol}$ is
infinite. In other words the central star stays above the same place
on the planet, as expected. Equation~(\ref{eq:theta3}) shows that
$\theta=0$ for all times in this limit, which is consistent with tidal
locking. Next apply Equation~(\ref{eq:theta3}) to the original,
idealized version of modern Earth. In this case $N_{sid}=361$ so that
\begin{equation}
\theta = 2\pi \frac{t}{(1440\ \min)}.
\end{equation}
Since $T_{sol}=1440$~min in this case, we have recovered our
definition of $\theta$.

\section{Results}
\label{sec:results}

\begin{figure}[h!]
\begin{center}
\epsfig{file=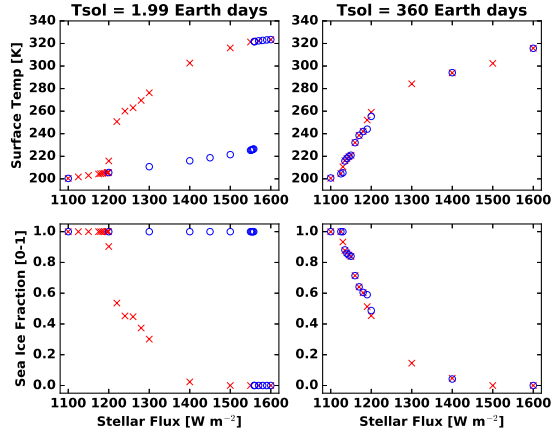, width=0.5\textwidth} 
\end{center}
\caption{Increasing the length of the solar day decreases the
  hysteresis associated with the snowball bifurcation. This plot shows
  the equilibrated global mean surface temperature (top) and sea ice
  fraction (bottom) for simulations with solar days of 1.99 Earth days
  (left) and 360 Earth days (right) as a function of the stellar flux.
  Red x's are warm start simulations and blue circles are cold start
  simulations.}
\label{fig:bifurcation_compare}
\end{figure}

For a solar day of 1.99 Earth days and an ocean mixed layer depth of
50~m, we reproduce the familiar Earth-like bifurcation diagram of
climate (Fig.~\ref{fig:bifurcation_compare}) with two separate climate
states that are bistable for a range of stellar flux as a result of
the ice-albedo feedback \citep{Budyko-1969:effect,Sellers-1969:ae}.
The ``warm'' state has large regions of open ocean and exists for
global mean surface temperatures above about 250~K. The ``snowball''
state is globally ice-covered and exists for global mean surface
temperatures below about 225~K. There is also a nearly, but not
completely, ice-covered state that occurs close to the transition to
global ice coverage (Fig.~\ref{fig:bifurcation_compare}). This is
similar to states discussed by \citet{Abbot-et-al-2011:Jormungand} and
\citet{rose2015stable}, but must be associated with a different
mechanism than those authors discussed since PlaSim uses the same
albedo for snow and bare sea ice and since we use a slab ocean with no
ocean heat transport. This state is associated with some hysteresis
and may be considered a distinct state, however, it does not appear in
most scenarios that we consider and we will simply consider it to be the
extreme branch of the warm state for the purposes of this paper.

When we increase the solar day to 360 Earth days, we find that the
bifurcations in global climate associated with the ice-albedo feedback
disappear (Fig.~\ref{fig:bifurcation_compare}), just as
\citet{checlair2017} found for the synchronously rotating case. For
this longer solar day the climate smoothly transitions to global ice
coverage. The ice-albedo feedback can still be seen through the
increase in the slope of the global mean surface temperature when the
planet is partially ice-covered (Fig.~\ref{fig:bifurcation_compare}),
but it is not strong enough to lead to hysteresis and bistability. In
a few cases some small differences between the equilibrated climate of
the warm and cold start simulations persist. \citet{checlair2017}
found that these small differences actually represent a continuum of
states allowed by the fact that the sea ice scheme only allows 100\%
ice-covered or ice-free gridboxes, and that they are not associated
with the standard, global ice-albedo feedback
\citep{Budyko-1969:effect,Sellers-1969:ae}. We can therefore conclude
that for a solar day of 360 Earth days the climate behaves effectively
as if the planet were tidally locked.

\begin{figure}[h!]
\begin{center}
\epsfig{file=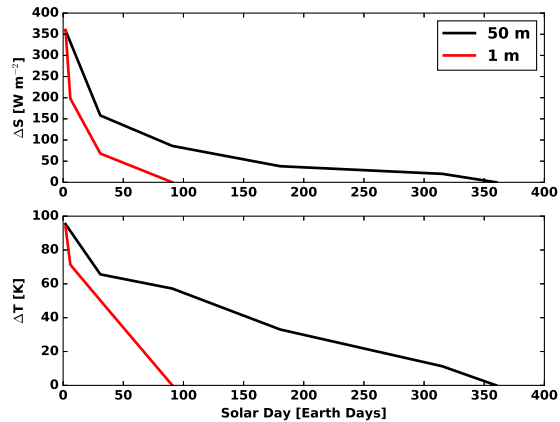, width=0.5\textwidth} 
\end{center}
\caption{This plot shows how the hysteresis and bistability are lost
  as the solar day is increased. Both the range of stellar flux over
  which the planet is bistable ($\Delta$S, top) and the increase in
  global mean surface temperature associated with the snowball-to-warm
  transition ($\Delta$T, bottom) are shown as a function of the solar
  day. See Fig.~\ref{fig:schematic} for a schematic diagram of
  $\Delta$S and $\Delta$T. Results are shown for simulations assuming
  the slab ocean has a mixed layer depth of 50~m (black) and 1~m
  (red).}
\label{fig:hysteresis}
\end{figure}

\begin{figure}[h!]
\begin{center}
\epsfig{file=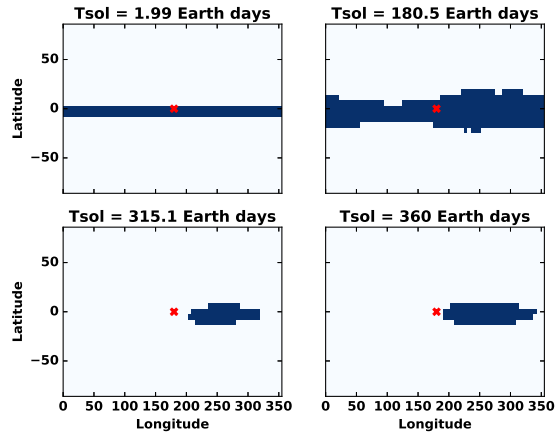, width=0.5\textwidth} 
\end{center}
\caption{As the solar day increases, the axis of symmetry changes.
  This plot shows maps of the sea ice coverage (very light blue) and
  open ocean coverage (dark blue) for climate states with a stellar
  flux just high enough to avoid global ice coverage for a variety of
  solar days. The red x represents the substellar point. For a solar
  day of 1.99 Earth days the surface ice pattern of the planet is
  zonally symmetric and for a solar day of 360 Earth days the surface
  ice pattern of the planet is symmetric around the axis connecting
  the substellar and anti-stellar points (with a slight eastward
  offset). For intermediate solar days the surface ice pattern of the
  planet transitions between these limits.}
\label{fig:sic_map}
\end{figure}

\begin{figure}[h!]
\begin{center}
\epsfig{file=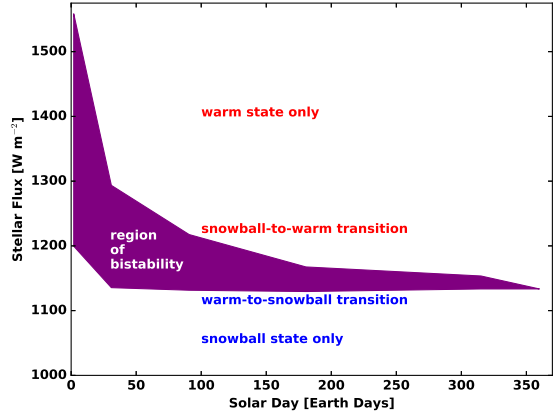, width=0.5\textwidth} 
\end{center}
\caption{This plot shows the stellar flux of the warm-to-snowball 
  transition and the snowball-to-warm transition as a function of
  the length of the solar day. The region of bistability between the
  bifurcations is colored magenta. The warm-to-snowball 
  transition is fairly independent of solar day, but the stellar flux
  of the snowball-to-warm transition decreases strongly as the
  solar day increases. A 50~m mixed layer depth is used in these
  simulations.}
\label{fig:hysteresis_fill1}
\end{figure}

\begin{figure}[h!]
\begin{center}
\epsfig{file=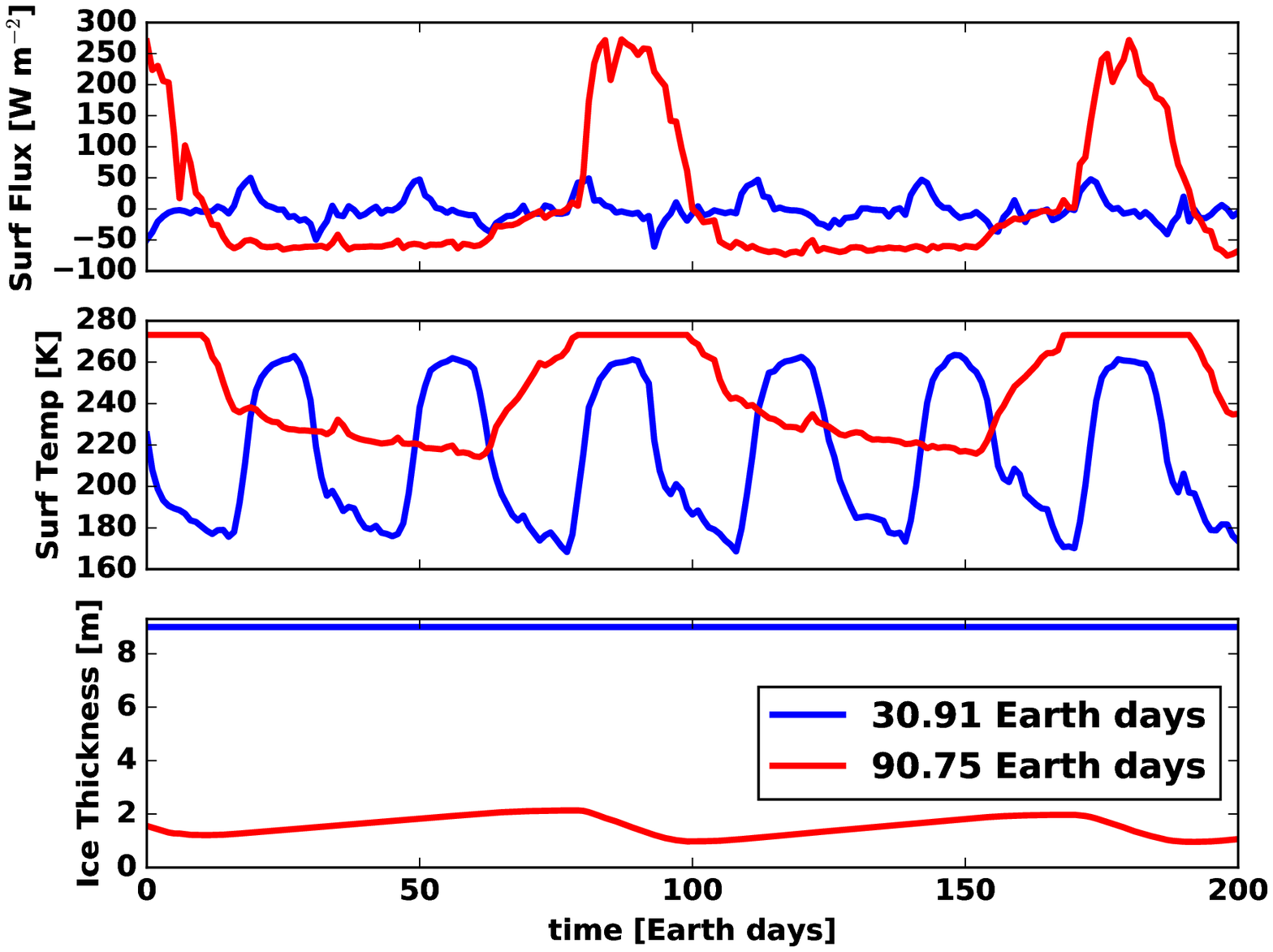, width=0.5\textwidth} 
\end{center}
\caption{This plot shows why increasing the length of the solar day causes the snowball-to-warm transition to occur at a lower stellar flux. The surface heat flux (top), surface temperature (middle), and ice thickness (bottom) for a gridbox at the equator are displayed as a function of time for solar days of 30.91 Earth days (blue) and 90.75 Earth days (red). The stellar flux is 1200~W~m$^{-2}$ and the model is in the snowball state in both cases.}
\label{fig:melt_out}
\end{figure}

As the solar day is increased, the amount of hysteresis drops rapidly
up to a solar day on the scale of 100 Earth days, and then slowly
until hysteresis is completely lost (Fig.~\ref{fig:hysteresis}). As
noted by \citet{boschi2013changes}, this loss of hysteresis is
accompanied by a general reorientation of the climate from symmetry
around the rotational axis (zonal symmetry) for small solar days to
something approaching the symmetry seen in synchronously rotating
planets (Fig.~\ref{fig:sic_map}). Synchronously rotating planets tend
to be roughly symmetric around the stellar flux axis that connects the
substellar and antistellar points
\citep{Pierrehumbert:2011p3287,koll-2015,koll-2016}. Maps of sea ice
for simulations with solar days of hundreds of Earth days have a
symmetry like this except that it lags the moving substellar point
slightly (Fig.~\ref{fig:sic_map}).

The loss of hysteresis as the solar day is increased is mostly
associated with a decrease in the stellar flux of the snowball-to-warm
transition (Fig.~\ref{fig:hysteresis_fill1}).
Figure~\ref{fig:melt_out} shows why the snowball-to-warm transition is
strongly affected by the length of the solar day. For longer solar
days the substellar point moves more slowly across the planet's
surface, allowing the temperature of the substellar region (at the
equator) to increase more (Fig.~\ref{fig:melt_out}). As a result,
melting occurs at the substellar point, which decreases the substellar
albedo due to melt pond formation, a parameterized process in PlaSim.
This leads to a dramatic increase in the stellar radiation absorbed by
the surface, which allows the equatorial regions to stay warm
throughout the year. As a result, the equatorial ice thins
(Fig.~\ref{fig:melt_out}). Once the ice becomes thinner than it's
seasonal cycle of about 2~m, it melts through to the ocean at the
substellar point, engaging strong ice-albedo feedbacks and causing
deglaciation from the snowball state.

In contrast, the stellar flux of the warm-to-snowball transition is
relatively constant as the length of the solar day is increased
(Fig.~\ref{fig:hysteresis_fill1}). As the solar day gets longer,
equatorial points experience larger diurnal cycles in surface
temperature. This makes the onset of glaciation during the night more
likely. For shorter solar days the planet has a zonal symmetry
(Fig.~\ref{fig:sic_map}), so that that nighttime glaciation is likely
to lead to global glaciation. For longer solar days the planet can
support solutions with ice during the night and open ocean during the
day at the equator (Fig.~\ref{fig:sic_map}). These states are often
associated with less day-side ice coverage than zonally symmetric
states at a similar stellar flux \citep{salameh2017role}. This means
that the stellar flux must be decreased more than would otherwise be
necessary in order to cause global glaciation for longer solar days.
The complex interaction between larger diurnal temperature cycles and
changes in planetary symmetry leads to a warm-to-snowball transition
that is nearly independent of the length of the solar day in PlaSim,
but this result may not be true in all models.

\begin{figure}[h!]
\begin{center}
\epsfig{file=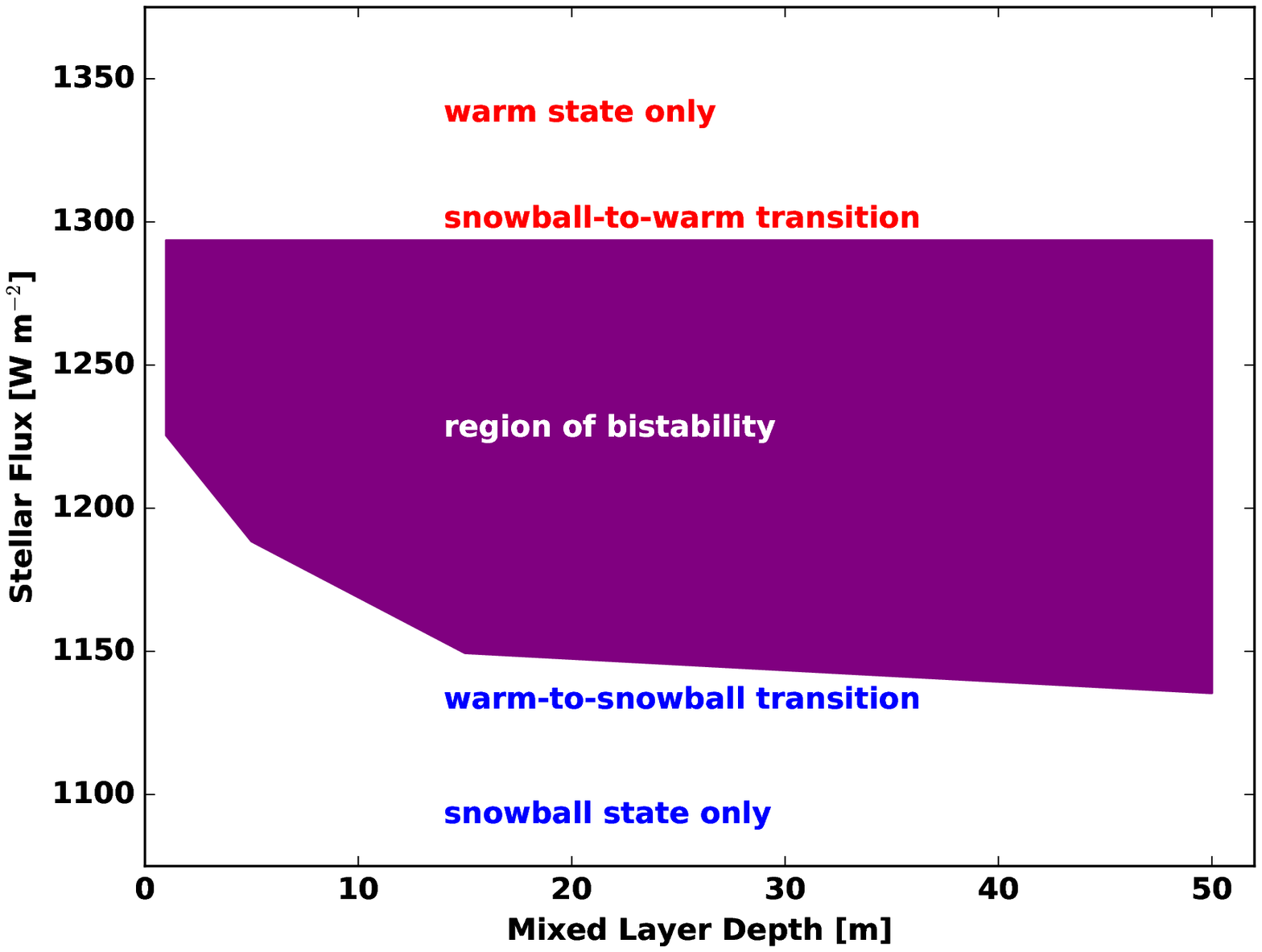, width=0.5\textwidth} 
\end{center}
\caption{This plot shows the stellar flux of the warm-to-snowball 
  transition and the snowball-to-warm transition as a function of
  the depth of the ocean mixed layer. The region of bistability
  between the bifurcations is colored magenta. The stellar flux of the
  snowball-to-warm transition is independent of the mixed layer
  depth, but the stellar flux of the warm-to-snowball transition
  decreases as the mixed layer depth increases. The solar day is 30.91
  Earth days in these simulations.}
\label{fig:hysteresis_fill2}
\end{figure}

\begin{figure}[h!]
\begin{center}
\epsfig{file=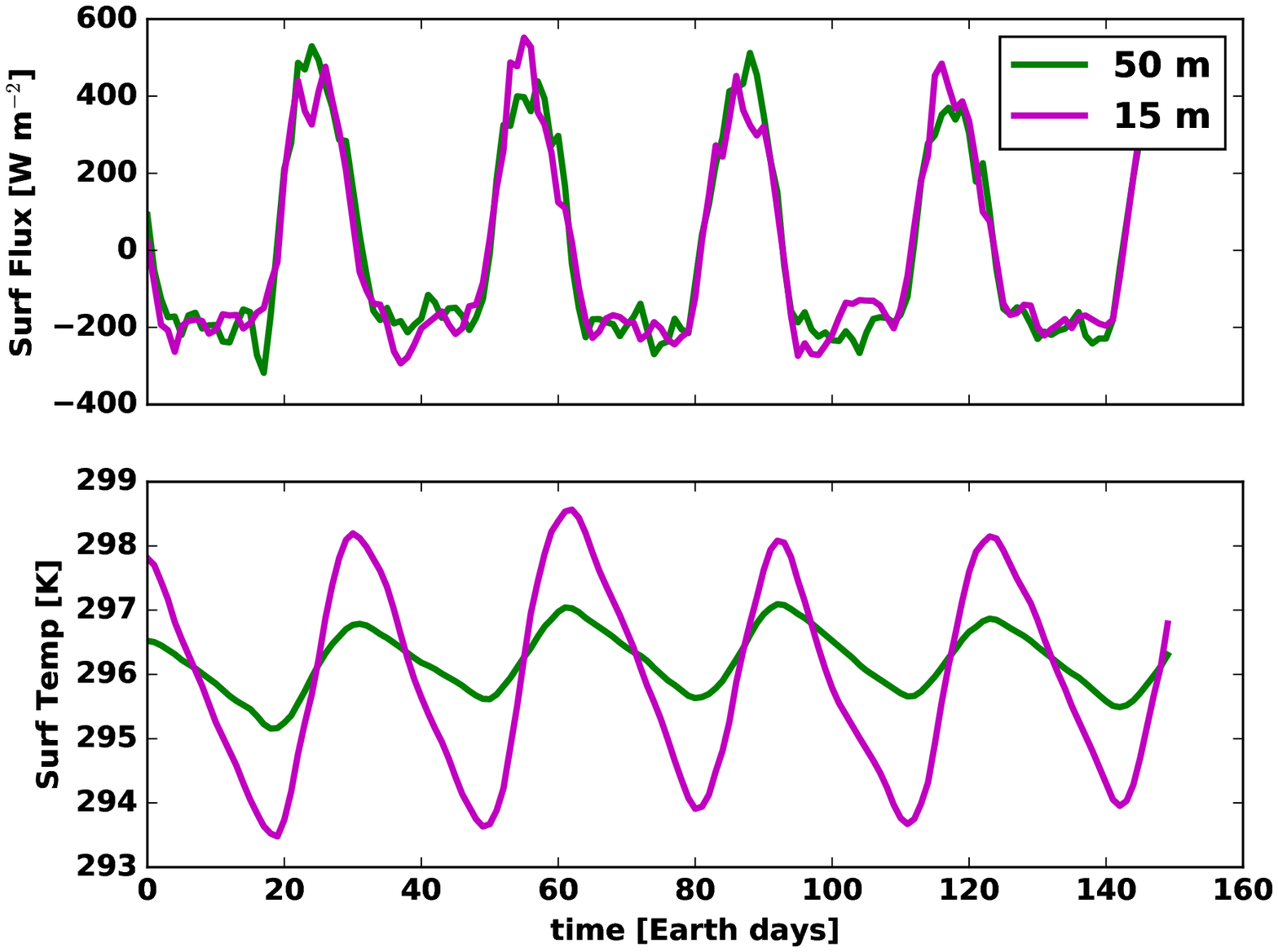, width=0.5\textwidth} 
\end{center}
\caption{This plot shows why increasing the ocean mixed layer depth
  causes the warm-to-snowball transition to occur at a higher
  stellar flux. The surface heat flux (top) and surface temperature
  (bottom) for a gridbox at the equator are displayed as a function of
  time for mixed layer depths of 50~m (green) and 15~m (magenta). The
  stellar flux is 1235~W~m$^{-2}$, the solar day is 30.91 Earth
  days, and the model is in the warm state in both cases.}
\label{fig:test_fluxes}
\end{figure}

Next we consider how changing the mixed layer depth affects the
snowball bifurcations and hysteresis. Using a mixed layer depth of 1~m
instead of 50~m significantly lowers the length of the solar day at
which hysteresis is lost (Fig.~\ref{fig:hysteresis}). Changing the
mixed layer depth does not affect the snowball-to-warm transition
(Fig.~\ref{fig:hysteresis_fill2}). The ocean is covered with ice when
this transition is encountered, so the ocean mixed layer depth does
not affect the surface climate. Decreasing the mixed layer depth
increases the stellar flux of the warm-to-snowball transition
(Fig.~\ref{fig:hysteresis_fill2}) in a straightforward way. All else
being equal, the main effect of decreasing the mixed layer depth is to
increase the magnitude of the surface temperature fluctuations along
the equator associated with the diurnal cycle in the warm state
(Fig.~\ref{fig:test_fluxes}). This means that for a smaller mixed
layer depth the minimum temperature in the cycle will cross the
freezing point, initiating the ice-albedo feedback that leads to
global glaciation, at a higher stellar flux. Consequently, the stellar
flux of the warm-to-snowball transition increases.

\section{Discussion}
\label{sec:discussion}

\citet{checlair2017} raised the question of how close a planet needs
to be to synchronous rotation in order to lose the snowball
bifurcations and climate hysteresis. Our results
(section~\ref{sec:results}) indicate that even slight increases in the
solar day (Fig.~\ref{fig:hysteresis}) significantly reduce climate
hysteresis, but the solar day needs to be hundreds of Earth days in
order to completely remove hysteresis. \citet{leconte2015asynchronous}
showed that planets in a circular orbit can get caught in spin states
with roughly two rotations per orbit. This means that their solar day
would be roughly equal to their orbital period
(section~\ref{sec:solarday}). Using the scalings in
\citet{kopparapu2016inner}, a planet that received the same stellar
flux as Earth would have an orbital period of 65 Earth days for a
stellar mass of 50\% of the Sun's. Our results suggest that a planet
with a solar day of 65 Earth days and a mixed layer depth of 50~m
would have hysteresis over about $\frac{1}{3}$ of the range of stellar
flux as modern Earth, and with a mixed layer depth of 1~m it would
have nearly completely lost hysteresis. If the stellar mass were 80\%
of the Sun's a similar planet would have a solar day of 209 Earth
days, making the reduction in hysteresis larger. Furthermore, these
reductions in hysteresis would be even more dramatic if the reduction
in the ice albedo due to the spectrum of a redder star were accounted
for \citep{Joshi:2012hu,shields2013effect,shields2014spectrum}.

This study represents an idealized investigation of the physical
mechanisms through which climate hysteresis is lost as the solar day
increases. For theoretical clarity, we have purposefully isolated
changes in the solar day while keeping other variables constant. To
apply these results to observed planets it will be necessary to
perform simulations using the best observational estimates of the
stellar spectrum, orbital distance, orbital period, eccentricity,
planetary rotation rate, planetary size, atmospheric mass, and all
other relevant variables. Accurately accounting for each of these
variables can effect the results, as \citet{kopparapu2016inner} showed
when they revisited the more idealized paper of \citet{yang2013}.

In this study we have not included ocean and thick ice dynamics. One
important aspect of ocean circulation is that on the night side of a
planet with a deep ocean, convection would be likely, extending the
mixed layer all the way to the bottom of the ocean. This would tend to
reduce the surface temperature drop experienced on the night side, and
delay the warm-to-snowball transition. Efficient flow of thick ice from
higher latitudes toward the equator
\citep{Goodman03,Pollard05,Goodman06,li2011sea,Tziperman2012,abbot13-snowball-circulation,pollard2017snowball}
would tend thicken the ice there, despite the mechanism described in
section~\ref{sec:results}. This would lessen the reduction in the
stellar flux of the snowball-to-warm transition as the length of the
solar day is increased. Both of these effects would tend to make the
hysteresis experienced by a planet larger than what we have estimated here.
Work with models including ocean dynamics and flowing thick ice is
required to determine the importance of these issues.

Many planets in the habitable zone with low CO$_2$ outgassing rates
may experience continuous cycling between warm and snowball
conditions, rather than equilibrating in the warm state
\citep{Tajika2007,Kadoya:2014kd,Menou2015,haqq2016limit,batalha2016climate,abbot-2016,paradise2017}.
The amount of time spent in the snowball state, which represents the
majority of the period of the cycles, is proportional to the CO$_2$
needed to cause deglaciation if we assume a constant CO$_2$ outgassing
flux \citep{abbot-2016}. Since increasing the solar day decreases the
amount of forcing necessary to cause snowball deglaciation
(Fig.~\ref{fig:hysteresis_fill2}), this means that planets with long
solar days and low CO$_2$ outgassing rates would experience much more
rapid cycles between the warm and snowball climate states. That said,
planets orbiting M and K-stars are less likely to experience climate
cycles at all due to the reduced ice-ocean albedo contrast
\citep{haqq2016limit}.

We have not taken into account the implications of CO$_2$ condensation
\citep{turbet2017co} in the snowball state in this paper. Since the
surface temperature can drop below 195~K, the sublimation point of
CO$_2$ at a surface pressure of 1~bar, even at the equator in the
snowball state (Fig.~\ref{fig:melt_out}), this is an issue that
warrants further investigation. CO$_2$ condensation would not
significantly affect the calculations we have made since we assumed a
low atmospheric CO$_2$ concentration and varied the stellar flux. It
could, however, be important in a snowball scenario involving the
carbon cycle and variations in CO$_2$
\citep{Walker-Hays-Kasting-1981:negative,Kirschvink92,Hoffman98}. The
slow-moving location of the substellar point and large diurnal
temperature variations, which would force CO$_2$ sublimation and
condensation, would make this problem particularly interesting.

The thermodynamic sea ice scheme in the model uses a 0D Semtner
formulation \citep{Semtner-1976:model}. This means there are no
interior levels in the ice, and that the temperature profile within
the ice is assumed to be linear between the melting point at the
bottom and the surface temperature at the top. This approximation will
yield reasonable heat fluxes through the ice and surface temperatures
as long as the ice is thinner than a few multiples of the critical
thickness, $z^\ast=\sqrt{\frac{\kappa P}{\pi}}$, where $P$ is the
period of variations in surface forcing (the solar day) and
$\kappa \approx 10^{-6}$ m$^2$ s$^{-1}$ is the thermal diffusivity of
ice \citep{Abbot-et-al-2010:semtner}. For $P$=30~Earth~days,
$z^\ast$=0.9~m and for $P$=90~Earth~days, $z^\ast$=1.6~m. This means
that the sea ice model should be providing a reasonable approximation
of surface temperature (and melting) near the equator when the ice is
thin enough (less than about 2~m) to lead to deglaciation from the
snowball state for most of the solar days considered here
(Fig.~\ref{fig:melt_out}). Nevertheless it might still be interesting
to investigate these processes using a more sophisticated sea ice
model.

A planet with an obliquity of 90$^\circ$ would experience variations
in stellar flux that are somewhat similar to a planet caught in a 2:1
spin-orbit resonance. Our results suggest that such a planet might
experience reduced hysteresis in planetary climate if its orbital
period is tens to hundreds of days. Calculations by
\citet{kilic2017multiple} using a similar modeling framework show that
this is indeed the case, especially when the warm state is defined as
any state with at least some open ocean, as we have done here.

\section{Conclusions}
\label{sec:conclusions}

We have used PlaSIM, an intermediate complexity global climate model
run with a mixed layer ocean and no ocean heat transport, to
investigate the snowball bifurcations and hysteresis in planetary
climate as a function of the length of the solar day. Our main
conclusions are as follows:

\begin{enumerate}
\item Hysteresis in planetary climate (the range in stellar flux for which
  there is bistability) is significantly reduced if the solar day is
  increased to tens of Earth days. This effect is stronger the smaller
  the mixed layer depth is. This would be relevant to any tidally
  influenced, but not synchronously rotating, planet orbiting an
  M or K-star.

\item Hysteresis in planetary climate is lost and the planet
  transitions smoothly into a snowball state if the solar day is
  increased to hundreds of Earth days. This could be relevant to
  tidally influenced planets orbiting late M and K-stars.

\item Increasing the length of the solar day mainly decreases
  hysteresis in planetary climate by decreasing the stellar flux of
  the snowball-to-warm transition. This happens because the substellar
  point is heated for longer if the solar day is longer, leading to
  melting and a reduction in albedo there, which leads to a higher
  mean surface temperature throughout the day and thinner ice. Once
  the ice is thin enough, it melts through to the ocean at its diurnal minimum,
  kicking off a large-scale ice-albedo feedback that leads to total
  deglaciation.

\item Decreasing the depth of the ocean mixed layer mainly decreases
  hysteresis in planetary climate by increasing the stellar flux of
  the warm-to-snowball transition. This is because all else being
  equal, a smaller mixed layer depth causes larger fluctuations in
  ocean surface temperature. This means that the diurnal minimum in
  equatorial surface temperature is more likely to cross the freezing
  point, leading to ice formation and a large-scale ice-albedo
  feedback that leads to complete glaciation.
\end{enumerate}

\section{Acknowledgements}
We thank R. Boschi for sharing and explaining his modifications to the
PlaSim radiation code and T. Komacek for comments on an early draft of
this paper. We acknowledge support from the NASA Astrobiology
Institute’s Virtual Planetary Laboratory, which is supported by NASA
under cooperative agreement NNH05ZDA001C. We acknowledge support from
NASA grant number NNX16AR85G, which is part of the ``Habitable
Worlds'' program. We acknowledge support from the National Science
Foundation under NSF award number 1623064. PP was supported by a NASA
Earth and Space Science Fellowship.


\end{document}